\documentclass{article}

\usepackage[preprint]{neurips_2025}
\usepackage[utf8]{inputenc}
\usepackage[T1]{fontenc}
\usepackage{hyperref}
\usepackage{url}
\usepackage{booktabs}
\usepackage{amsfonts}
\usepackage{amsmath}
\usepackage{graphicx}
\usepackage{xcolor}
\usepackage{microtype}
\usepackage{enumitem}
\usepackage{array}

\title{Trajectory Supervision for Continual Tool-Use Learning in LLMs}

\author{%
  Vishnu Vardhan Reddy B \\
  CS 590NN \\
  UMass Amherst \\
  \texttt{vbheemreddy@umass.edu} \\
  \And
  Sagnik Chatterjee \\
  CS 590NN \\
  UMass Amherst \\
  \texttt{sagnikchatte@umass.edu} \\
  \And
  Soumik Bhatta \\
  CS 590NN \\
  UMass Amherst \\
  \texttt{sbhatta@umass.edu} \\
}

\begin{document}

\maketitle

\begin{abstract}
Most language-model training data shows final artifacts, not the process that produced them. We study a tractable version of this question in tool use: when a model learns a stream of new API domains, does keeping tool-use trajectories help compared with stripping the intermediate API trace? We fine-tune Llama 3.1 8B Instruct with QLoRA on API-Bank using four sequential domain blocks. Condition A strips previous API request/response lines from the prompt and trains the model to predict the next API call. Condition B keeps the trajectory context. In a single-seed pilot, full held-out generation evaluation shows that Condition B reaches 56.9\% final exact full-call accuracy compared with 39.2\% for Condition A. B also improves final API-name accuracy by 7.7 points. However, B uses 25.1\% more training tokens, the run uses one seed, and the task is next-call prediction rather than full dialogue success.
\end{abstract}

\section{Introduction}

Language models are usually trained on finished text: answers, articles, code, and other final products. The intermediate process that produced the answer is usually missing. Process-supervision work in reasoning addresses this \citep{lightman2023verify}, but it is hard to test directly because real human thought traces are private, expensive, and often unavailable.

Tool-use data gives a narrower but concrete proxy. A tool-use example can include a user request, an API call, the API response, and the next assistant action. These traces are not human internal reasoning. They are external interaction records. Still, they expose an action-observation sequence that final-response data removes. Our project asks whether this extra trajectory signal changes how a model adapts when tool domains arrive sequentially.

We use continual learning as a stress test, not as a new algorithmic contribution. If two supervision formats perform similarly on one domain, differences may become clearer when the model learns a sequence of new domains and must retain older ones. We therefore split API-Bank \citep{li2023apibank} into four sequential blocks and evaluate after each training block on all blocks.

We compare a stripped-context next-API-call baseline against a trajectory-context baseline under the same model, seed, stream ordering, QLoRA setup, and evaluation code. We focus on whether retaining API trajectory context changes continual next-call learning.

Our main findings are:
\begin{itemize}[leftmargin=*]
    \item Full held-out generation evaluation favors trajectory context. After training on all four blocks, Condition B reaches 56.9\% exact full-call accuracy, compared with 39.2\% for Condition A.
    \item The largest improvement is in selecting the right tool. B reduces final wrong-API errors from 102 to 12, while exact parameter generation remains imperfect.
    \item The evidence is limited by one seed and a token-budget confound. B trains on 2.32M tokens versus 1.86M for A, so the result supports a hypothesis but does not isolate the cause.
\end{itemize}

\section{Related work}

\paragraph{Process supervision.}
Lightman et al.\ compare process and outcome supervision for mathematical reasoning and show that step-level supervision can be more effective than final-answer supervision \citep{lightman2023verify}. We use that distinction, but our data is different. API trajectories are not annotated reasoning steps; they are records of tool calls and tool observations. They are less direct evidence about human reasoning, but easier to collect in real systems.

\paragraph{Tool-use learning.}
Toolformer showed that language models can learn to call external tools from self-supervised tool-use traces \citep{schick2023toolformer}. API-Bank provides a compact benchmark for tool-augmented language models with API calls, responses, and dialogue context \citep{li2023apibank}. ToolLLM and ToolBench scale this idea to many real-world APIs and solution paths \citep{qin2023toolllm}. We use API-Bank because it is small enough for repeated QLoRA experiments in a course project and still has structured tool-call supervision.

\paragraph{Continual learning in LLMs.}
Sequential fine-tuning can damage previously learned behavior. TRACE formalizes continual learning evaluation for LLMs and reports severe forgetting under sequential training \citep{wang2023trace}. We use the same evaluation setup: after training on block $D_i$, evaluate on all blocks $D_j$. Here, the question is whether the supervision format itself changes adaptation and retention, without adding replay or regularization.

\paragraph{Efficient adaptation.}
LoRA trains low-rank adapter weights instead of updating all parameters \citep{hu2022lora}. QLoRA adds 4-bit quantization to make fine-tuning larger models feasible on limited hardware \citep{dettmers2023qlora}. We use QLoRA with Llama 3.1 8B Instruct \citep{dubey2024llama} so that both conditions can be trained and checkpointed in Colab.

\section{Method}

\subsection{Task and data}

We use API-Bank and construct a stream of four domain blocks, $D_1$ through $D_4$. Each block contains a disjoint set of API examples. Training proceeds sequentially: train on $D_1$, evaluate on all blocks, then train on $D_2$, and so on. This produces a $4 \times 4$ evaluation matrix. Rows are training stage and columns are evaluation block.

The full held-out generation evaluation scores only examples whose expected output contains a parseable API call. The final-stage scored evaluation totals are 126, 104, 103, and 107 examples for blocks $D_1$ through $D_4$, respectively. The training notebooks also run a faster sampled evaluation with 32 examples per block after each stage.

\subsection{Conditions}

Condition A is the stripped-context baseline. It removes prior \texttt{API-Request} and \texttt{API-Response} lines from the input context before training and generation evaluation. It still predicts the next API call when the target output is an API call. This is narrower than the original proposal's broad ``outcome-only'' framing, so we describe it as stripped-context next-call supervision.

Condition B is the trajectory-context condition. It keeps the previous API request and response lines in the prompt. This gives the model access to the action-observation sequence before the next API call. Both conditions use the same base model, stream ordering, seed, optimizer settings, and scoring code.

\subsection{Model and training}

All runs use \texttt{meta-llama/Llama-3.1-8B-Instruct}. We fine-tune with QLoRA using 4-bit NF4 quantization, LoRA rank 32, LoRA alpha 64, bfloat16 computation, maximum sequence length 1024, and an effective batch size of 16. We use AdamW with a learning rate of $2 \times 10^{-4}$. Each block is trained for three epochs. The training runs use seed 42.

The main fairness limitation is token count. Condition A consumes 1,857,169 training tokens; Condition B consumes 2,324,314 training tokens. B therefore sees 25.1\% more tokens because trajectory context is longer. Token-matched training is left as future work.

\subsection{Evaluation metrics}

For generation, we greedily decode the model's next action and parse API calls with a regular expression of the form \texttt{[ApiName(param='value')]}. We report:
\begin{itemize}[leftmargin=*]
    \item \textbf{API-name accuracy}: the generated API name matches the expected API name.
    \item \textbf{Exact full-call accuracy}: the generated API name and normalized parameter dictionary exactly match the expected call.
    \item \textbf{Name-plus-any-param accuracy}: the API name is correct and at least one expected parameter-value pair is correct.
    \item \textbf{Malformed/no-call rate}: the generated text does not contain a parseable API call.
\end{itemize}

For the sampled continual-learning analysis, we also compute average accuracy (AA), backward transfer (BWT), forward transfer (FWT), average forgetting, and area under the learning curve (AULC) using exact full-call accuracy.

\section{Results}

\subsection{Sampled continual-learning evaluation}

Table~\ref{tab:sampled} summarizes the 32-example-per-block evaluation produced during training. B has higher final average accuracy and higher forward transfer, but it also has more negative BWT and larger average forgetting under this sampled metric. So B improves final AA and FWT, but the sampled run still shows forgetting on earlier blocks.

\begin{table}[t]
\centering
\caption{Sampled generation evaluation from the training notebook. Scores are exact full-call accuracy unless noted.}
\label{tab:sampled}
\resizebox{\linewidth}{!}{%
\begin{tabular}{lrrrrrr}
\toprule
Condition & Final AA & BWT & FWT & Avg. forgetting & AULC & Train tokens \\
\midrule
A: stripped context & 38.3 & -10.4 & 22.9 & 10.4 & 41.8 & 1.86M \\
B: trajectory context & 53.9 & -13.5 & 33.3 & 13.5 & 57.2 & 2.32M \\
\bottomrule
\end{tabular}}
\end{table}

\begin{figure}[t]
\centering
\includegraphics[width=\linewidth]{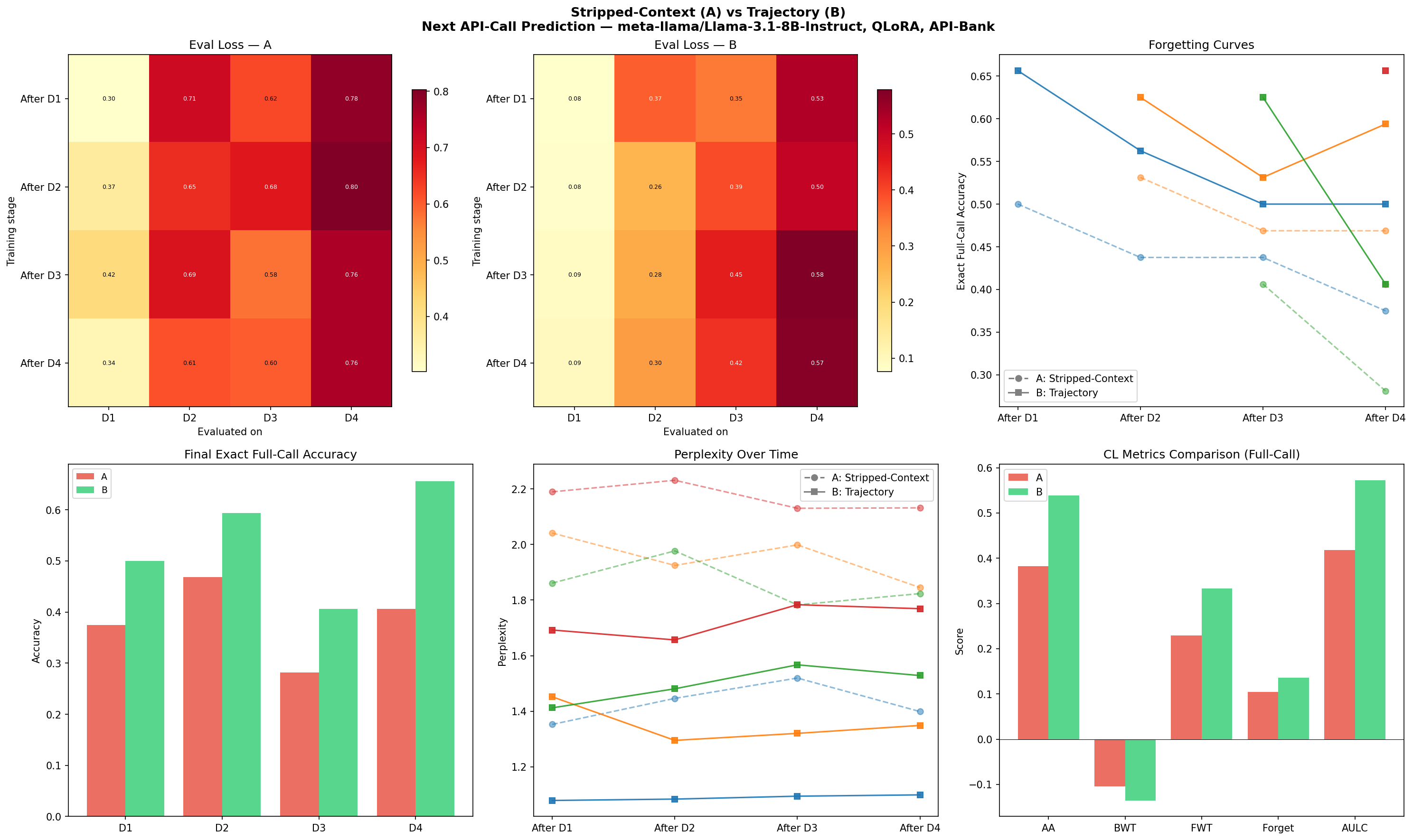}
\caption{Sampled evaluation used during training. Condition B obtains lower loss and higher full-call accuracy in the sampled analysis, while both conditions show nonzero forgetting.}
\label{fig:sampled}
\end{figure}

\subsection{Full held-out generation evaluation}

\begin{figure}[t]
\centering
\includegraphics[width=0.92\linewidth]{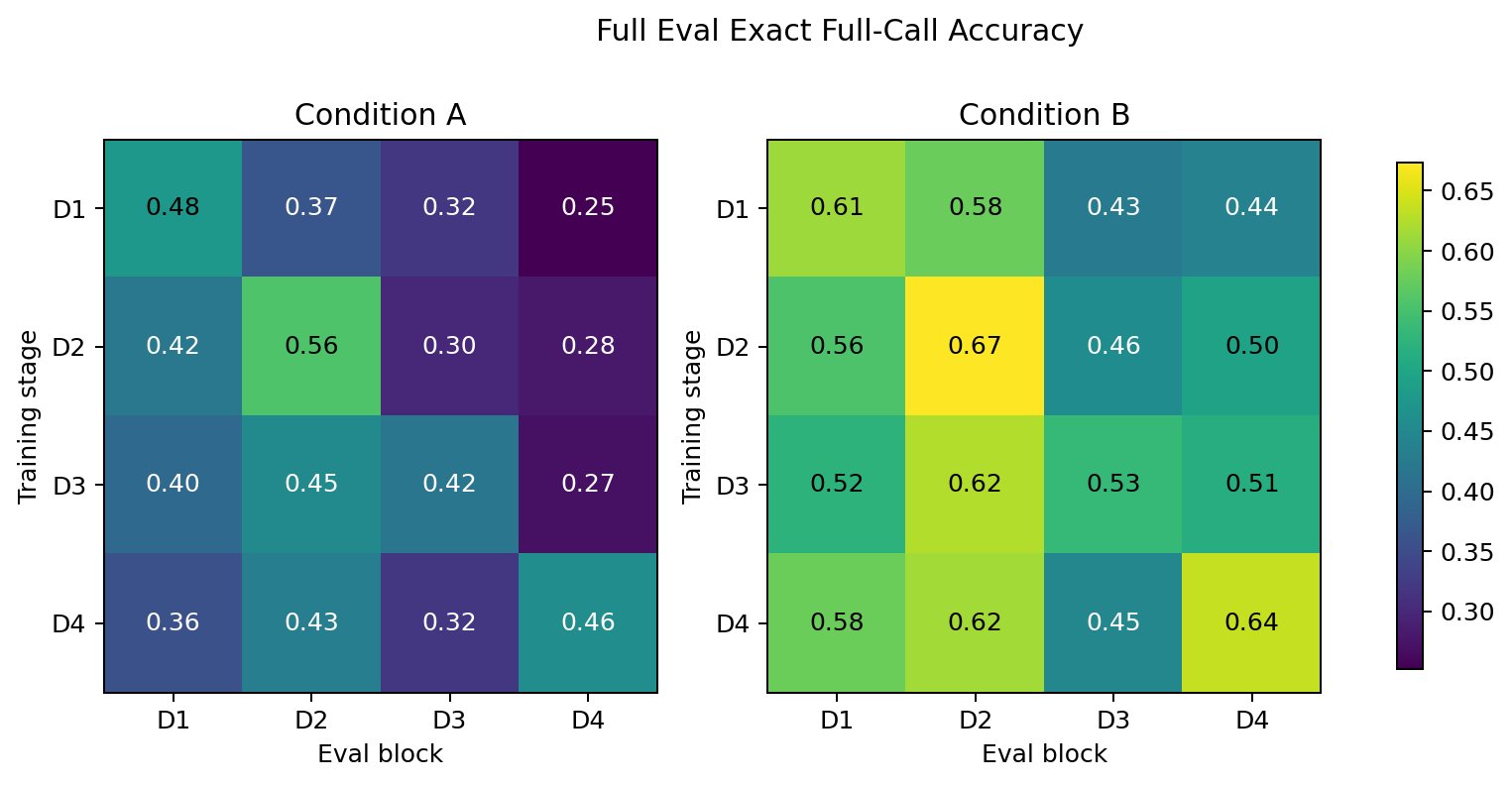}
\caption{Full held-out exact full-call accuracy. Rows are training stages and columns are evaluation blocks. Condition B is higher on every final-stage block and most earlier stages.}
\label{fig:fullheat}
\end{figure}

After training, we ran a separate full generation evaluation over every scored held-out example for each saved adapter. This removes the 32-example sampling limit used during training. Figure~\ref{fig:fullheat} shows the exact full-call accuracy matrices.

\begin{figure}[t]
\centering
\includegraphics[width=0.92\linewidth]{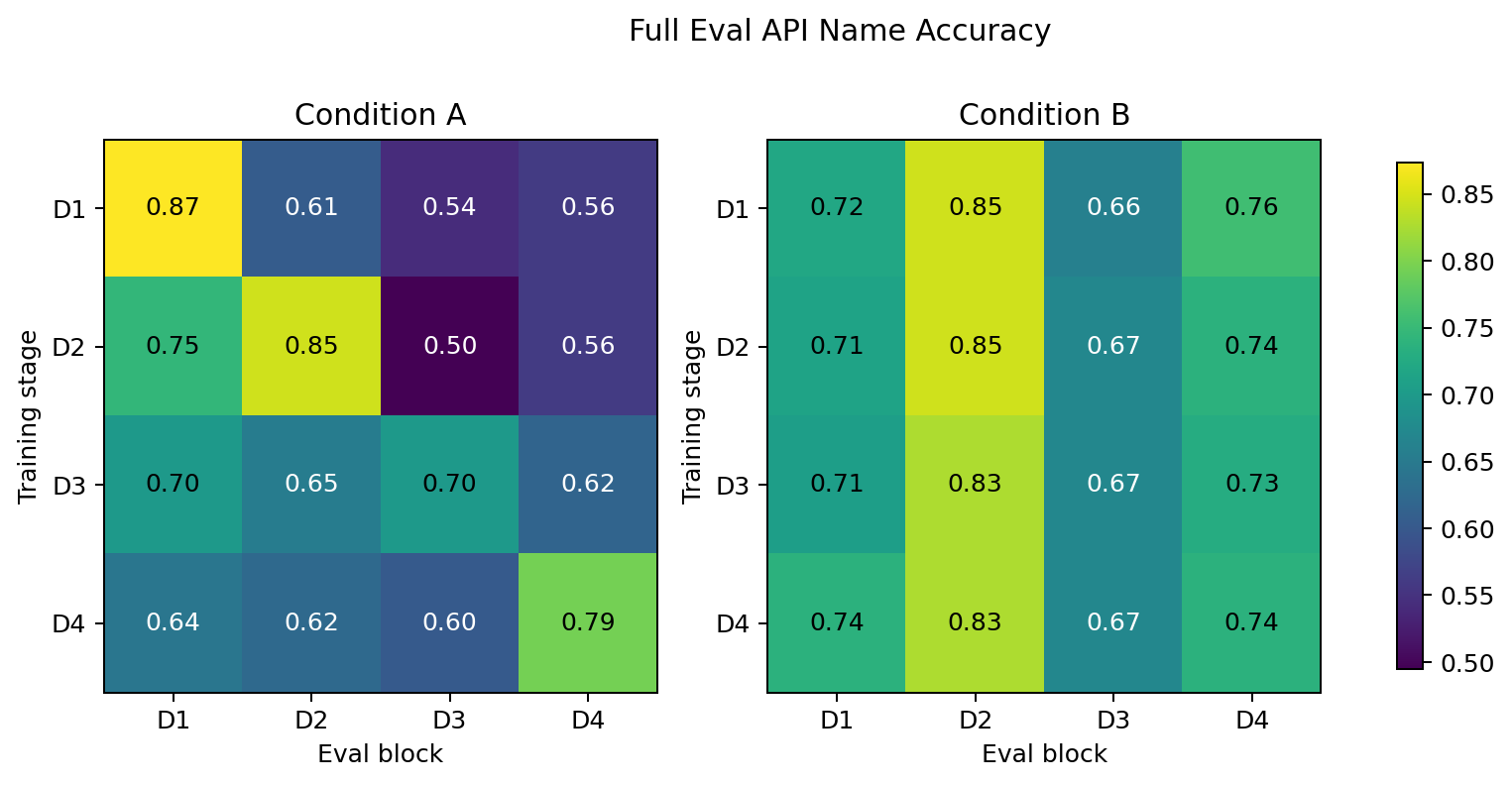}
\caption{Full held-out API-name accuracy across all training stages and evaluation blocks. Condition A shows high diagonal values (0.87 at D1/D1, 0.85 at D2/D2) but drops sharply off-diagonal, indicating forgetting. Condition B maintains more uniform accuracy across all blocks, with values clustered in the 0.71--0.85 range.}
\label{fig:nameheat}
\end{figure}

The final-stage comparison is the clearest result. Condition B reaches 56.9\% mean exact full-call accuracy, compared with 39.2\% for Condition A, a difference of 17.7 percentage points. On API-name accuracy, B reaches 74.3\% compared with 66.6\% for A, a difference of 7.7 points. Table~\ref{tab:fullfinal} shows the final-stage block-level results.

\begin{table}[t]
\centering
\caption{Full held-out final-stage generation results after training through $D_4$.}
\label{tab:fullfinal}
\resizebox{\linewidth}{!}{%
\begin{tabular}{llrrrrr}
\toprule
Condition & Metric & D1 & D2 & D3 & D4 & Mean \\
\midrule
A & Exact full-call & 35.7 & 43.3 & 32.0 & 45.8 & 39.2 \\
B & Exact full-call & 57.9 & 61.5 & 44.7 & 63.6 & 56.9 \\
\midrule
A & API-name & 64.3 & 62.5 & 60.2 & 79.4 & 66.6 \\
B & API-name & 73.8 & 82.7 & 67.0 & 73.8 & 74.3 \\
\midrule
A & Name + any param & 51.6 & 56.7 & 50.5 & 65.4 & 56.1 \\
B & Name + any param & 67.5 & 76.9 & 61.2 & 72.0 & 69.4 \\
\bottomrule
\end{tabular}}
\end{table}

\begin{figure}[t]
\centering
\includegraphics[width=0.82\linewidth]{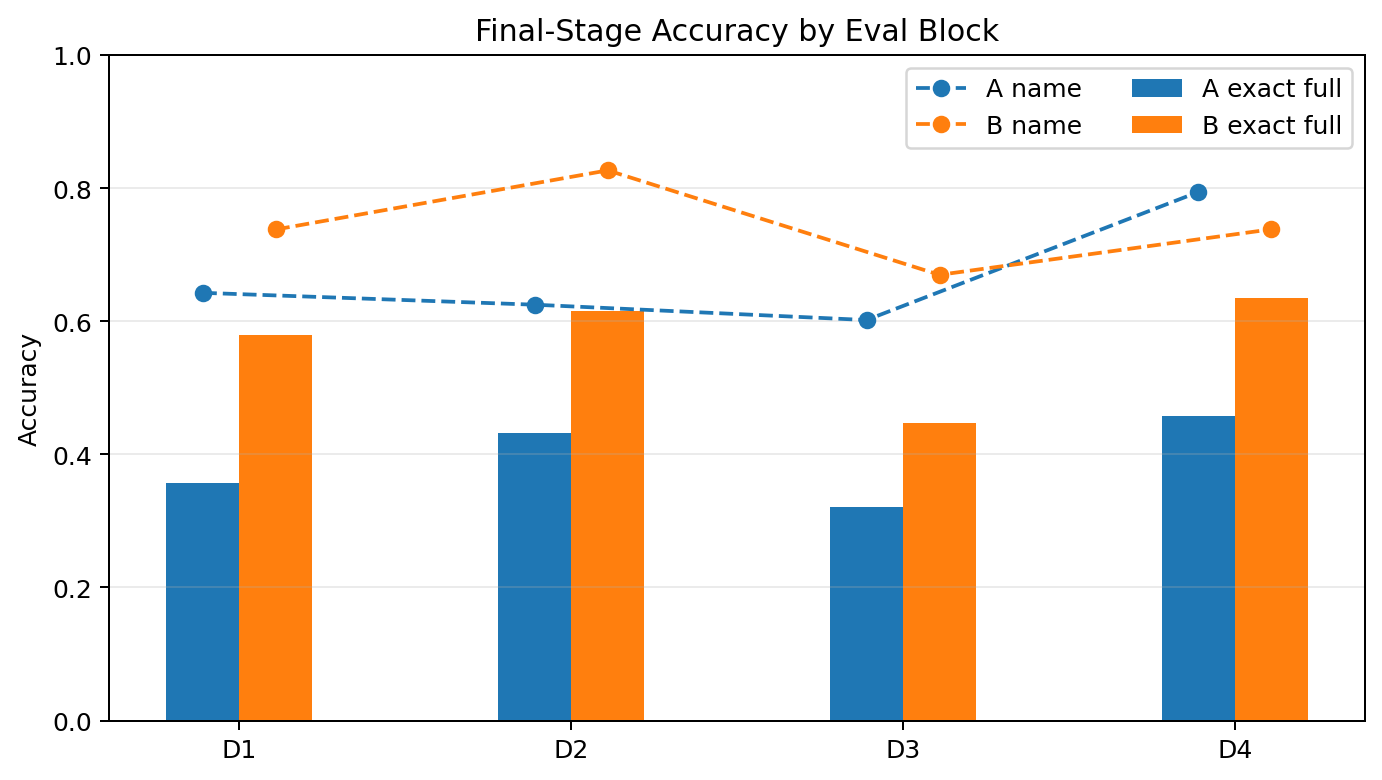}
\caption{Final-stage full evaluation by block. B improves exact full-call and name-plus-any-param accuracy on every block.}
\label{fig:finalblock}
\end{figure}

\subsection{Error analysis}

The error categories clarify what trajectory context changed. At the final stage, A produced 102 wrong-API errors, while B produced only 12. This suggests trajectory context helps the model pick the correct tool. However, B also produced 101 malformed/no-call errors, compared with 45 for A. The B model often had the right tool structure at the name level but did not always produce an exactly parseable or parameter-complete call.

\begin{table}[t]
\centering
\caption{Final-stage full-eval error category counts across all scored blocks.}
\label{tab:errors}
\begin{tabular}{lrr}
\toprule
Category & A & B \\
\midrule
Exact full call & 172 & 251 \\
Correct API, some params & 74 & 54 \\
Correct API, wrong params & 47 & 22 \\
Wrong API & 102 & 12 \\
Malformed or no call & 45 & 101 \\
\bottomrule
\end{tabular}
\end{table}

\section{Discussion}

In this single-seed API-Bank stream, keeping trajectory context improves final held-out next-API-call generation. The improvement is strongest for tool selection. The action-observation history likely helps the model pick which API to call next.

This result does not prove that process supervision is inherently better. The most obvious confound is token budget: B sees 25\% more tokens, and more data can help even if the extra content is not causally important. We also have only one seed, so we cannot report confidence intervals. API-Bank traces are structured and partly synthetic, which makes them a weaker stand-in for real reasoning data. And our task is next API-call prediction, not full end-to-end task success, which is a narrower evaluation than what a deployment setting would require.

B makes fewer wrong tool choices, but it makes more parse failures. Longer trajectory prompts may help the model identify the right tool while making exact output formatting harder to get right. Our parser is also strict: a call with the correct API and nearly correct parameters still counts as a failure if the string does not exactly match. A semantic parameter scorer alongside exact-match scoring would help separate these effects.

\section{Conclusion and future work}

We tested whether trajectory context changes continual tool-use learning in a controlled Llama 3.1 8B QLoRA pilot. The full held-out evaluation favored trajectory context on final exact API-call generation: 56.9\% versus 39.2\%. The comparison is still limited by the larger token count for B and by the single seed.

The next step is to run multiple seeds and report confidence intervals. A token-matched variant of Condition B would also help separate trajectory content from simply seeing more tokens. Longer task streams and a semantic parameter scorer would make the evaluation stronger. Replay or retrieval methods would be interesting, but they would be new continual-learning interventions rather than part of the A/B supervision question.

\section{Contributions}

Table~\ref{tab:contrib} lists individual responsibilities. All three of us contributed equally. We discussed the setup, interpreted the results, and edited the report.

\begin{table}[h]
\centering
\caption{Project contributions.}
\label{tab:contrib}
\small
\setlength{\tabcolsep}{4pt}
\begin{tabular}{p{0.27\linewidth}p{0.65\linewidth}}
\toprule
Student & Contributions \\
\midrule
Vishnu Vardhan Reddy B & Data preprocessing pipeline, evaluation and scoring code, checked generated outputs, full generation evaluation runs. \\
Sagnik Chatterjee & Training notebook implementation, continual-learning metric computation, Colab experiment execution. \\
Soumik Bhatta & Error analysis, figure generation, presentation preparation, checked error categories and experiment outputs. \\
\bottomrule
\end{tabular}
\end{table}

\end{document}